\documentclass[12pt]{article}

\usepackage[english]{babel}

\usepackage{amssymb}        	
\usepackage{mathtools}      	
\usepackage{bm}         		
\usepackage{url}                
\usepackage[misc]{ifsym}        

\newcommand{\bfa}       	{{\bf a}}
\newcommand{\bfb}       	{{\bf b}}
\newcommand{\bfc}       	{{\bf c}}
\newcommand{\bfv}       	{{\bf v}}
\newcommand{\bfx}       	{{\bf x}}
\newcommand{\bfy}       	{{\bf y}}
\newcommand{\bfz}       	{{\bf z}}
\newcommand{\bfA}       	{{\bf A}}
\newcommand{\bfB}       	{{\bf B}}
\newcommand{\bfC}       	{{\bf C}}
\newcommand{\bfG}         {{\bf G}}
\newcommand{\bfM}         {{\bf M}}
\newcommand{\bfX}       	{{\bf X}}

\newcommand{\bbR}        {{\mathbb R}}
\newcommand{\bbC}        {{\mathbb C}}

\newcommand{\bfgamma}{{\boldsymbol{\gamma}}}
\newcommand{\g}             	{\gamma}

\newcommand{\up}        	{\uparrow}
\newcommand{\dn}        	{\downarrow}
\newcommand{\Pauli}     	{\bm{\sigma}}
\newcommand{\bfone}      {{\bf 1}}

\newcommand{\pair}[2]   	{\left(\begin{array}{c} \!\! #1 \!\! \\ \!\! #2 \!\! \end{array}\right)}
\newcommand{\row}[2]    	{\begin{array}{c} \big({\;#1^*\ \ #2^*\;}\big) \\ \phantom{x} \end{array}\!\!}
\newcommand{\mat}[4]    	{\left({\begin{array}{cc} #1 &\ \ #2 \\ #3 &\ \ #4\end{array}}\right)}
\newcommand{\quat}[4]   	{\left({\begin{array}{c} #1 \\  #2 \\ #3 \\ #4\end{array}}\right)}
\newcommand{\minus}         {\!\!-}
\newcommand{\tall}[1]   	{{\displaystyle \mathop{ #1 }_{\phantom{o}}^{\phantom{o}}}}

\newcommand{\diceone}       {\lower2pt\hbox{\Cube{1}}}
\newcommand{\dicetwo}       {\lower2pt\hbox{\Cube{2}}}
\newcommand{\dicethree}     {\lower2pt\hbox{\Cube{3}}}
\newcommand{\dicefour}      {\lower2pt\hbox{\Cube{4}}}
\newcommand{\dicefive}      {\lower2pt\hbox{\Cube{5}}}
\newcommand{\dicesix}       {\lower2pt\hbox{\Cube{6}}}

\begin{document}
\title{The Arithmetic of Uncertainty unifies Quantum Formalism and Relativistic Spacetime}


\author
{John Skilling,$^{1}$ Kevin H. Knuth,$^{2\ast}$\\
\normalsize{$^{1}$Maximum Entropy Data Consultants Ltd, Kenmare, Ireland}\\
\normalsize{$^{2}$University at Albany (SUNY) Albany NY 12222, USA}\\
\normalsize{$^\ast$Corresponding author; E-mail:  kknuth@albany.edu.}
}

\date{}

\baselineskip24pt

\maketitle

\begin{abstract}
The theories of quantum mechanics and relativity dramatically altered our understanding of the universe ushering in the era of modern physics.
Quantum theory deals with objects probabilistically at small scales, whereas relativity deals classically with motion in space and time.

We show here that the mathematical structures of quantum theory and of relativity follow together from pure thought, defined and uniquely constrained by the same elementary ``combining and sequencing'' symmetries that underlie standard arithmetic and probability.
The key is uncertainty, which inevitably accompanies observation of quantity and imposes the use of \emph{pairs} of numbers.

The symmetries then lead directly to the use of complex ``$\surd\mathord-1$'' arithmetic, the standard calculus of quantum mechanics, and the Lorentz transformations of relativistic spacetime.
One dimension of time and three dimensions of space are thus \emph{derived} as the profound and inevitable framework of physics.
\end{abstract}

\section{Introduction}

Science relies on numeric quantification, which can be traced back to Euclid, Galileo, and Newton.
While many now take this success for granted, some scientists have pondered the ``unreasonable effectiveness of mathematics'' \cite{Wigner:1960,Hamming:1980} as well as its necessity \cite{Feynman:1967},
and others continue to strive to understand the foundations of the quantum mechanical formalism \cite{Miller:2012,Timpson:2013,Hardy:2016,Friedberg+Hohenberg:2018,Caticha:2019,Jaeger:2019}.
Such questions are deep.
For example, we quantify speed with a single real number, a scalar.
But we must quantify velocity with three real numbers.
Why?  And how do we know?

This paper is about engineering the formalism of quantification \cite{Pfanzagl:1968}.
To accomplish this, we use the basic symmetries beneath combination and sequencing, which underpin formal science.
Only such supremely elementary assumptions have such sufficient range and authority to be an acceptable base for the fundamental language of wide-ranging science.
Requiring some subtle detail (continuity perhaps) would be less compelling because it could more easily be denied.
Robust foundation needs to be simple.
Hoping for wide accessibility, our account eschews excessive detail and is intended to be accessible to neophyte undergraduate students of reasonable diligence and moderate numeracy.

Classical objects might optimistically be measurable to arbitrary precision: ``\emph{Given any precision requirement, there could exist technology to accomplish it}''.
That would mean that quantity could be treated as a single number, ignoring any intrinsic but negligible uncertainty.
But this is a double limit and the pessimistic view ``\emph{Given any technological ability, there will always exist requirements that defeat it}'' could also be held.
Classically, it's not clear whether ``$\textit{requirement}\rightarrow 0$'' or ``$\textit{technology}\rightarrow 0$''
 should dominate, which leaves the status of single-number quantities uncomfortably in question.

Fortunately there is no ambiguity in the quantum world, where an elementary target can not be measured with precision by some arbitrarily smaller probe because there are no such probes.
It becomes impossible to bootstrap complete knowledge of a target because we start correspondingly ignorant about any probe that we might use.
It follows that a faithful description of a quantum target requires a \emph{pair} of numbers representing a fusion of quantity and uncertainty.
The connection will be more intimate than just ``quantity $\pm$ error bar'', which would really be just a conventional couple of scalars.

The basic symmetries impose a specific calculus on number pairs, in keeping with but more subtle than standard scalar arithmetic.
We derive \emph{complex arithmetic} operating on pairs which we recognise as quantum amplitudes, observable through modulus-squared probabilities.
Not only do we construct the Feynman picture of quantum mechanics, but we find that these same symmetries also lead to the Pauli matrices which generate spin, energy and momentum, and beyond that to 3+1-dimensional relativistic spacetime.
The physics of quantity-with-uncertainty is to be described within this required mathematical formalism.

This adopts the strategy of \cite{Cox:1946} to develop mathematical language in accordance with the relevant fundamental symmetries (here, of physics).
With this perspective, it is not surprising that mathematics works as the language of physics \cite{Knuth:FQXI2015}.
The mathematics we use works because it is engineered to work.

Our guiding principle is parsimony of laws.
Any operation that our mathematics does allow should be allowed unless prevented by some new law, and that means that the mathematics must be as simple as we can make it.
If we occasionally stray into technical language, it is not with intent to claim erudition or sophistication, but rather to reach out to those of greater erudition and sophistication than ourselves while aiming to explain to wider readership why the framework of physics must be as it is.

\section{Addition and Multiplication}

Sum and product rules are the foundation of arithmetic and thence of the rich structure of mathematics that science uses to model the physical world.

We assume that separate objects can exist.
Although we illustrate this here with spots on the faces of dice, we do not attempt to define the nature of the objects in question.
Applications are legion and we do not place limits on the objects or properties that users might have in mind.
We just assume {\bf commutativity} (order doesn't matter for the purpose in hand) illustrated with
\begin{equation}
    \underbrace{\; \dicethree \textrm{\ with\ } \dicefive \,=\, \dicefive \textrm{\ with\ } \dicethree\;}_\textrm{commutative, $a\mathord+b=b\mathord+a$}
\end{equation}
and {\bf associativity} (brackets don't matter either)
\begin{equation}
    \underbrace{ \big(\,\dicethree \textrm{\ with\ } \dicefive\,\big) \textrm{\ with\ } \diceone \,=\, \dicethree \textrm{\ with\ } \big(\,\dicefive \textrm{\ with\ } \diceone\,\big) }_\textrm{associative, $(a\mathord+b)\mathord+c = a\mathord+(b\mathord+c)$}
\end{equation}
This \underline{associative commutativity} implies that the mathematical representation of quantification is additive \cite{Aczel:associativity:2004,Jaynes:Book,Knuth+Skilling:2012} (up to isomorphism, which allows changing the labels while preserving the content).
Usage of addition is the \emph{sum rule}, here obtained in a way that will upgrade into quantum theory.

We also assume that what can be added up can be replicated, subject to left- and right-{\bf distributivity} (replication applies to any target)
\begin{equation}
    \underbrace{
     \begin{array}{r} 4 \textrm{\ of\ } \big(\,\dicethree \textrm{\ with\ } \dicefive\,\big) = \big(\,4 \textrm{\ of\ } \dicethree\,\big) \textrm{\ with\ } \big(\,4 \textrm{\ of\ } \dicefive\,\big)   \\[5pt]
                              ( 3 \textrm{\ with\ } 5 )  \textrm{\ of\ }  \dicefour \ = \big(\,3 \textrm{\ of\ } \dicefour\,\big) \textrm{\ with\ } \big(\,5 \textrm{\ of\ } \dicefour\,\big)   \end{array}
      }_\textrm{distributive, $a(b\mathord+c) = ab\mathord+ac$ and $(a\mathord+b)c = ac\mathord+bc$}
\end{equation}
and non-trivial {\bf associativity} (replications can be chained without forced annihilation)
\begin{equation}
   \underbrace{\; 3 \textrm{\ of\ } \big(\, 4 \textrm{\ of\ } 2 \textrm{\ of\ }\big) \,\diceone \,= \big(\,3 \textrm{\ of\ } 4 \textrm{\ of} \,\big) \,2 \textrm{\ of\ } \diceone\;}_\textrm{associative,   $a(bc) = (ab)c$}
\end{equation}
This \underline{associative distributivity} implies that its representation is multiplicative (up to change of units, for consistency with addition).
Partitioning is the inverse, where proportions multiply down instead of replicates multiplying up \cite{Smith&Erickson:associativity:1990,Caticha:2009:rational-belief,Knuth:laws,Knuth:measuring,Knuth+Skilling:2012}.
Usage of multiplication is the \emph{product rule}.

Associative commutativity and associative distributivity are our \emph{foundational symmetries}.\\		
{\textbf{Nothing else is needed.}}
Children use these informally as they learn about addition and multiplication through shuffling and grouping.
The supreme simplicity of these ideas tokens the generality of application that we need for the deepest basis of science.
This explains the success of mathematical modelling of a world in which separate objects can exist (commutativity) and which can behave independently of others (distributivity).
The symmetries force arithmetical rules to which we can have no alternative.
The content of our modelling is up to us, but the language (in this case standard arithmetic) is defined.

One can think of the mathematics used in science as having been engineered to be consistent with these symmetries, thus ensuring its success.
It is then no mystery that mathematics works \cite{Wigner:1960,Hamming:1980}, because it could not have been any other way \cite{Knuth:FQXI2015}.
The uniqueness (up to isomorphism  \cite{Knuth:measuring,Knuth+Skilling:2012,Skilling+Knuth:MPQ}) and extreme familiarity of the rules appear to give them independent status, whereas in fact they were deliberately engineered.
\textbf{If the foundational symmetries are accepted, the rules are forced, and the resulting mathematics becomes the quantitative language of physics.}

Specifically, measure theory applies the sum rule to quantification, with additivity being the unique formalism for ubiquitous situations.
Probability calculus applies the sum and product rules to partitioning of allowed possibilities \cite{Knuth+Skilling:2012,Skilling+Knuth:MPQ}, which allows us to learn about the world by eliminating some of what was previously deemed possible.
Again, probability is the unique calculus in ubiquitous situations.
The content of the calculus is up to us, but the language is forced.	

There are many applications of the rules, in human affairs as well as in science.
For example, money is an application of measure leading to betting as a subsidiary application of probability \cite{DeFinetti:1931:original,Caves:2000}.
But applications are not foundations.

Where objects obeying these symmetries have only one relevant property, the standard representation is scalar.
However, objects may have several properties.
For example, six-sided dice have individuality (1 per die) and can display different numbers of spots (ranging from 1 to 6).	
Those properties, the number of dice thrown and the number of displayed dots, are separately additive.

Multi\-dimensional addition is straightforwardly comp\-onentwise.
Multi\-dimensional multiplication, though, is not quite uniquely defined by the founding symmetries.
Distributivity requires that a product is bilinear (linear in each factor) but associativity does not altogether remove the remaining ambiguity.
Investigating this leads simply and directly to the basic language of physics in the form of quantum formalism and relativistic spacetime.

\section{Quantum Foundation}

Our world is quantised.
There is an irreducible quantum of targets of a given type, and there are likewise irreducible investigative probes.
Our knowledge always derives from interactions, which means that we can never attain complete knowledge of any individual object, whether it be a target investigated with an incompletely known probe, or a probe interacting with an incompletely known target.
Consequently, our knowledge of objects will always be accompanied by inherent uncertainty.
The connection between quantity and uncertainty is potentially more intimate than just ``quantity $\pm$ error bar'', which
means that \textbf{our description needs to fuse quantity with uncertainty into  two-parameter ``pairs''} that we write as
\begin{equation}
	\bfx = \binom{x_1}{x_2}, \quad \bfy = \binom{y_1}{y_2}, \;\; \dots
	\end{equation}
Precisely how quantity and uncertainty are to be encoded by these pairs is to be determined.

\underline{Associative commutativity}
\begin{equation}
    \underbrace{\bfx + \bfy = \bfy + \bfx}_\textrm{commutative} \qquad\textrm{and}\qquad \underbrace{(\bfa + \bfb) + \bfc = \bfa + (\bfb + \bfc)}_\textrm{associative}
\end{equation}
implies the sum rule
\begin{equation}
    	\bfx + \bfy = \pair{x_1}{x_2} + \pair{y_1}{y_2} = \pair{x_1 + y_1}{x_2 + y_2}
\label{eq:pairadd}
\end{equation}
now in two-dimensional pair-wise form.

Upon interaction, \underline{associative distributivity} then gives
\begin{equation}
    \underbrace{
     \begin{array}{r} \bfa\cdot(\bfb+\bfc) = \bfa\cdot\bfb + \bfa\cdot\bfc   \\
                              (\bfa + \bfb) \cdot \bfc =  \bfa\cdot\bfc + \bfb\cdot\bfc  \end{array}
      }_\textrm{distributive, left and right}
      \quad\textrm{and}\quad
 	\underbrace{(\bfa \cdot \bfb) \cdot \bfc = \bfa \cdot (\bfb \cdot \bfc)}_\textrm{associative}
\end{equation}
because summation has to remain linear regardless of probing or targeting context (distributivity) and probing or targeting is a sequential process (associative).
Left distributivity ensures that a pair product is linear in the second factor and right distributivity ensures linearity in the first, so that  the multiplication is bilinear, taking the form
\begin{equation}
	(\bfx\cdot\bfy)_i = \sum_{j,k=1}^2 \gamma_{ijk} x_j y_k
\label{eq:gamma}
\end{equation}
where the $\gamma$'s are eight constants which take standard values in standardised coordinates.
\textbf{This is why physics is fundamentally linear.}
Probing with a pair $\bfx$ applies a linear $2\mathord\times2$ matrix
\begin{equation}
	(\bfx\cdot)_{ik}= \sum_{j=1}^2 \gamma_{ijk} x_j
\label{eq:probe}
\end{equation}
to the target pair $\bfy$.

Unlike for scalars, pair multiplication is not unique.
We encounter similarly nonunique multiplication through dot and cross products in vector calculus (though those products are not full-rank 3-vectors).
Here, we seek multiplication in which the $\bbR^2\mathord\times\bbR^2$ pair products remain full-rank pairs in $\bbR^2$.
We find \emph{three} settings for the $\gamma$'s in which products \hbox{$\bfx\cdot\bfy$} retain full non-degenerate pair status.
(Details are in the Appendix.)
They represent classes \emph{not} related to each other by any real coordinate transformation.
Taking standard coordinates, these are	
\begin{equation}
\begin{split}
	&\pair{x_1}{x_2}\cdot\pair{y_1}{y_2} =													\\
	&\		\underbrace{\pair{x_1y_1 - x_2y_2}{x_1y_2 + x_2y_1}}_\textrm{A}    		\textrm{\ \ or\ \ }
			\underbrace{\pair{x_1y_1 + x_2y_2}{x_1y_2 + x_2y_1}}_\textrm{B}    	\textrm{\ \ or\ \ }
			\underbrace{\pair{x_1y_1}{x_1y_2 + x_2y_1}}_\textrm{C}
\end{split}
\label{eq:complex}
\end{equation}
Probing with $\bfx$ can thus be specified by any of the three $2\mathord\times2$ matrices
\begin{equation}
    (\bfx\, \cdot) = 	\underbrace{\mat{x_1}{\minus x_2}{x_2}{x_1}}_\textrm{A}    	\textrm{\quad  or\quad}
                			\underbrace{\mat{x_1}{x_2}{x_2}{x_1}}_\textrm{B}           		 \textrm{\quad  or\quad}
                			\underbrace{\mat{x_1}{0}{x_2}{x_1}}_\textrm{C}
\label{eq:pairmul}
\end{equation}
These varied possibilities allow the richness of physics while limiting the possibilities to those allowed by A and B and C alone.

Lest repeated operation of $(\bfx\,\cdot)$ cause targets $\bfy$ to diverge towards infinity or collapse towards zero, thereby exploding or imploding them, we use $\det(\bfx\,\cdot) = 1$, which defines unit quantity of $\bfx$.	
There is no loss of generality because $(\bfx\,\cdot)$ can always be rescaled in any particular case.
These normalised operators have only one free parameter $\phi$ related to $x_2/x_1$ and take the form
\begin{equation}
    (\bfx\, \cdot)  =	\underbrace{\mat{\!\!\cos\phi}{\minus \sin\phi\!\!}{\!\!\sin\phi}{\cos\phi\!\!}}_\textrm{A}         	\textrm{\  \ or\ \ }
                 		\underbrace{\mat{\!\!\cosh\phi}{\sinh\phi\!\!}{\!\!\sinh\phi}{\cosh\phi\!\!} }_\textrm{B}      	\textrm{\  \ or\ \ }
                 		\underbrace{\mat{1}{0}{\phi}{1}}_\textrm{C}
\label{eq:phi}
\end{equation}
We can also focus on the nature of the multiplication itself by building $\phi$ from infinitesimal increments of operators
\begin{equation}
	\bfA = \mat{0}{\minus 1}{1}{0}  \textrm{\quad or\quad } \bfB = \mat{0}{1}{1}{0} \textrm{\quad or\quad}  \bfC = \mat{0}{0}{1}{0}
\label{eq:ABC}
\end{equation}
so that
\begin{equation}
	(\bfx\,\cdot) = \lim_{n\rightarrow\infty} \left( \bfone + \frac{\phi}{n}\bfG \right)^n = \exp(\phi \bfG)
\label{eq:generator}
\end{equation}
where the \emph{generator} $\bfG$ is the matrix $\bfA$ or $\bfB$ or $\bfC$.

The first of these (operator A) is rotation by phase $\phi$.
Taking (\ref{eq:pairadd}) for addition and (\ref{eq:pairmul}A) for multiplication, we recognise the sum and product rules of complex arithmetic, so that pairs are complex numbers
\begin{equation} \label{eq:complex-number}
    \pair{x_1}{x_2} = re^{i\phi}
\end{equation}
obeying the standard rules.
\ Moreover, unit quantity \linebreak is identified with unit determinant, which for (\ref{eq:pairmul}A) is modulus-squared $\det(\bfx\,\cdot) = |\bfx|^2   =1$.
\textbf{Hence the inherent uncertainty in a unit object refers to what remains undefined in $\bfx$, namely phase $\phi$, so that each new object brings with it an unknown phase.}
At this point, \emph{probability} must enter the development.


\section{Probability}

Our ignorance of phase is uniformly distributed, giving phase a uniform probability distribution
\begin{equation}
    \Pr(\phi) = \frac{1}{2\pi}\,,\quad (0 \le \phi < 2\pi)
\end{equation}
around the circle.
Otherwise, we could use identically-formed components to build composite objects $\bfX = \bfx_1 + \bfx_2 + \dots + \bfx_n$ whose overall phase would be arbitrarily precise around the supposed mode, thus defeating the engineering requirement of inevitable uncertainty.
Meanwhile, phase is necessarily continuous because any restriction would conflict with general application of the complex sum rule.
Continuity is not an assumption, it's a requirement.

We have no way of extracting what might be a definitive ``truth''.
The best we can do in the face of uncertainty is use replication to obtain predictions of average behaviour, with precision increasing statistically as the square root of the replication factor.
These average predictions are called \emph{ensemble averages}.

According to the standard rules of probability, ensemble averaging involves summing (integrating) over all the unknown parameters.
Such averaging is known in statistics as ``marginalization'' by analogy with the treatment of tables of possibilities laid out on a page.
In quantum theory here, we average over uniformly distributed phases to get
\begin{equation}
\begin{split}
	\Big\langle |\bfX|^2 \Big\rangle &= \Big\langle |\bfx_1 + \bfx_2 + \dots + \bfx_n|^2 \Big\rangle_{\phi_1,\phi_2,\dots,\phi_n} \\&= |\bfx_1|^2 + |\bfx_2|^2 + \dots + |\bfx_n|^2.
\end{split}
\label{eq:phaseaverage}
\end{equation}
Hence by summing over the unknown phases, the quantity we have access to via experiment is the probability, or likelihood, of a given outcome.
\textbf{This is why quantum mechanics is a probabilistic theory.}
This likelihood, which is additive because of the scalar sum rule, is found to refer to the squared modulus, which is the Born rule \cite{Born:1926}.

It remains to try the alternative operators B and C.
Taking (\ref{eq:pairmul}B) or (\ref{eq:pairmul}C) for multiplication, we would again identify unit quantity with unit determinant.
Uncertainty would again refer to $\phi$ --- now a pseudo-phase --- and ignorance of pseudo-phase would again be uniformly distributed.
\begin{equation}
	\Pr(\phi) = \textrm{constant}\,,\quad (-\infty < \phi < \infty)
\label{eq:improper}
\end{equation}
However, probabilities sum to unity and infinite range makes this normalising constant zero.
Such calculus would be unusable because any finite range of $\phi$ would be infinitely improbable.
The engineering would have failed through the distribution (\ref{eq:improper}) being improper, leaving operator A as the only possibility.

This foundation of quantum theory is elementary and simple.
Quantity and uncertainty fuse together into complex numbers with uncertainty referring to phase (Fig.~1)
and modulus-squared being the observable quantity (the Born rule for arbitrary amount) representing ensemble averages.

\begin{figure}[ht]	
\begin{center}
	\includegraphics[width=180pt]{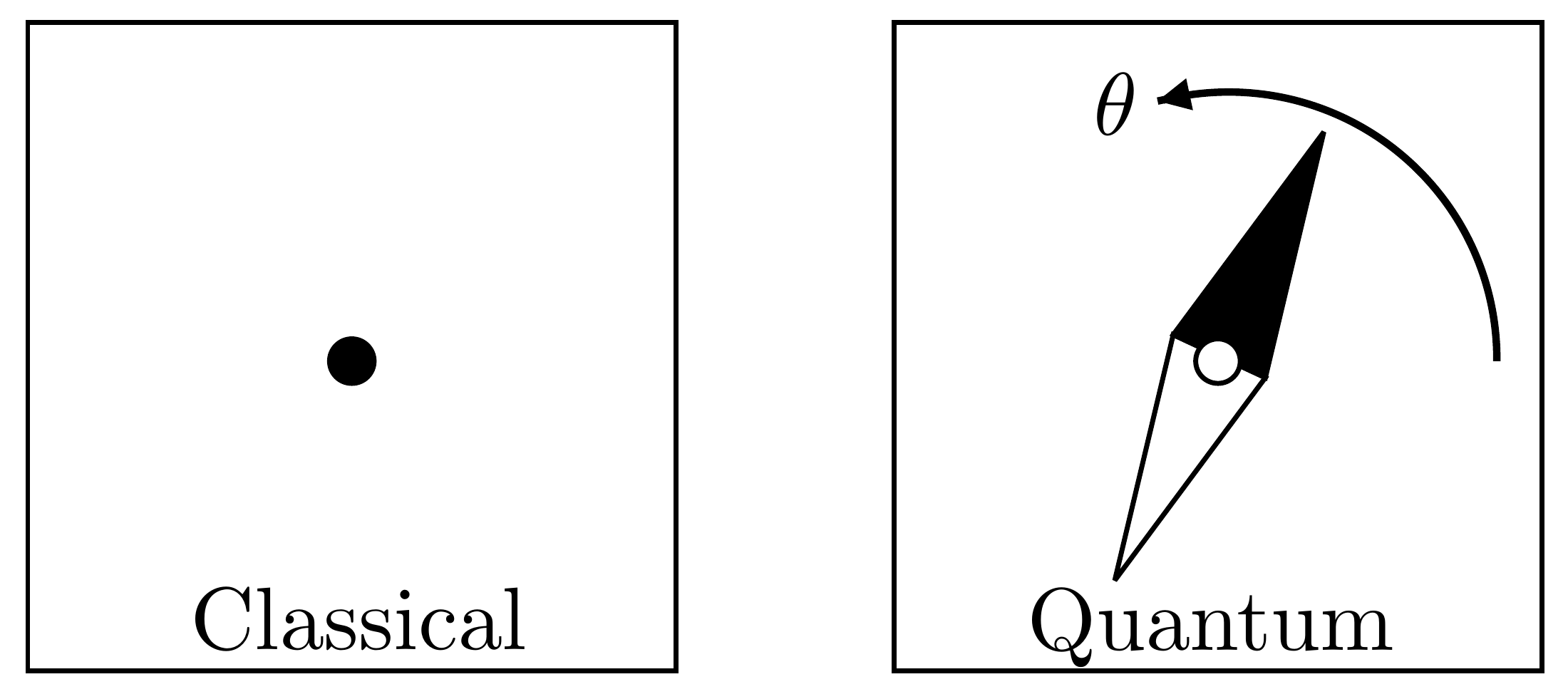}
\end{center}
\caption{Classical and Quantum objects}
\end{figure}

\section{Inference}

Looking ahead and anticipating spacetime which has not yet been developed, physical apparatus produces beams of particles emitted sporadically at some average rate, with intensity determined by averaging over unknown phases.
Particles themselves will be secondary to beams and the predictions we make are probabilistic, in the form of an ensemble of possibilities.
Standard probability, often redundantly called ``Bayesian'',  affords a solid foundation for the traditional quantum formalism.

The probabilistic nature of quantum theory has been acknowledged by  \cite{Caves+Fuchs+Schack:2002,Fuchs:2002,Pitowsky:2003,Bub:2007,Fuchs:2010qbism}
 in ``Quantum Bayesianism'' (QBism), though it's actually intrinsic and does not need to be argued or demonstrated separately.
Indeed, our formalism was engineered to be automatically consistent with probability theory through obeying the same foundational symmetries \cite{Knuth+Skilling:2012,Skilling+Knuth:MPQ}.
Contradiction was never possible, so extra assumption was never needed.

Quantum formalism is part of physics: it predicts the likely (probabilistic) behavior of specified models of physical situations.
Given those likelihoods, Bayesian analysis then computes posterior probabilities, which assess the models in the light of outcomes as actually observed.
But Bayes' Theorem, which formalises our inferences about the world, is not a part of physics.
Physics makes predictions quantified probabilistically in terms of likelihoods.
The physical world proceeds independently of our thoughts about it (except insofar as we may decide to interrupt its natural evolution).
We recommend keeping physics and inference conceptually apart \cite{Jaynes:mysteries}.

\section{Qubits}

So far, we have investigated the calculus of objects whose only relevant property is existence (in some identified state).
We now upgrade to objects that can exist in two states, $\uparrow$ and $\downarrow$ say, which we can produce by combining independent $\up$ and $\dn$ ensembles.
Instead of having a single-state ``quantity-with-uncertainty'' representation with two real numbers fused into a single complex, we now have a representation involving two complex (four real) numbers.
\begin{equation}
	\psi = \pair{\psi_\up}{0} + \pair{0}{\psi_\dn} = \pair{\psi_\up}{\psi_\dn} = \pair{\psi_0\mathord+i\psi_1}{\psi_2\mathord+i\psi_3}
\label{eq:psi}
\end{equation}
Physicists identify such objects as spin-half leptons while mathematicians call them 2-spinors of rank 1.
Taking our cue from computation, we call them ``qubits'' for short.
Any number of states greater than two can be decomposed in binary fashion, so using just two loses no generality.

Quantification of  $\up$ and $\dn$ separately is performed with detectors which are now represented by $2\mathord\times2$ projection matrices
\begin{equation}
\begin{split}
    q_\up &= |\psi_\up|^2 = \row{\psi_\up}{\psi_\dn} \mat{1}{0}{0}{0} \pair{\psi_\up}{\psi_\dn} \\
    q_\dn &= |\psi_\dn|^2 = \row{\psi_\up}{\psi_\dn} \mat{0}{0}{0}{1} \pair{\psi_\up}{\psi_\dn}
\end{split}
\end{equation}
Under change of basis and linear combination, this generalises to $q = \psi^\dagger \bfM \psi$ for general quantification of a qubit where $\bfM$ is an arbitrary matrix
 --- Hermitian by construction and because any anti-Hermitian part would cancel.
The observable ensemble average is $\langle q \rangle  =  \langle \psi^\dagger \bfM \psi\rangle$, commonly expressed as
  $\langle q \rangle = \textrm{trace}(\rho\bfM)$ where $\rho = \langle \psi \psi^\dagger \rangle$ is the \emph{density matrix} though we do not use that construction here.

Under combination, qubits still obey associative distributivity so that their representation is two-dimensional multiplicative as before,
except that it's over the complex field instead of the real.
We could now be allowed all three of A and B and C, now used as complex-to-complex $\bbC^2\mathord\times\bbC^2\rightarrow\bbC^2$ operators.
In this expanded context there is no reason to exclude B or C, but that is the limit of the allowed freedom.
Technically, $\bfA$ and $\bfB$ become inter-convertible by complex transformation but $\bfC$ retains its individuality.
This implies that $\bfC$ will provide a quantification distinct from that provided by $\bfA$ and $\bfB$.

Note that $\phi$ in (\ref{eq:phi}) can also be complex now, so that the matrices $\bfA,\bfB,\bfC$ can have complex coefficients.
Leaving $\bfC$ aside for now,  $\bfA$ and $\bfB$ yield the Pauli matrices \cite[pp. 164-165]{Sakurai:1994}
\begin{equation}
    \Pauli_x = \underbrace{\mat{0\ }{1}{1\ }{0}}_\bfB               	\,,\qquad
    \Pauli_y = \underbrace{\mat{0}{\minus i}{i}{0}}_{i\bfA}         	\,,\qquad
    \Pauli_z = \underbrace{\mat{1}{0}{0}{\minus 1}}_{\bfB\bfA}
\end{equation}
which have been written in Hermitian form for convenience and which (together with the identity matrix $\Pauli_0=\bfone$) form a closed group under multiplication.

The three Pauli matrices yield quantification
\begin{equation}
    \begin{array}{ll}
         q_x = \psi^\dagger \Pauli_x \,\psi = 2(\psi_0\psi_2 + \psi_1\psi_3)                	& = 2\, {\tt Re}(\psi_\up^* \psi_\dn) 		\\[4pt]
         q_y = \psi^\dagger \Pauli_y \,\psi = 2(\psi_0\psi_3 - \psi_1\psi_2)                	& = 2\, {\tt Im}(\psi_\up^* \psi_\dn) 		\\[4pt]
         q_z = \psi^\dagger \Pauli_z \,\psi = \psi_0^2 + \psi_1^2 - \psi_2^2 - \psi_3^2     	& = |\psi_\up|^2 - |\psi_\dn|^2       		\end{array}
\label{eq:rates}
\end{equation}
The overall quantity
\begin{equation}
    \begin{array}{ll}
         q_0 = \psi^\dagger\, \bfone \,\psi = \psi_0^2 + \psi_1^2 + \psi_2^2 + \psi_3^2       & = |\psi_\up|^2 + |\psi_\dn|^2       \end{array}
\label{eq:quantity}
\end{equation}
is not independent, the relationship being
\begin{equation}
    q_x^2 + q_y^2 + q_z^2 = q_0^2
\label{eq:pure}
\end{equation}
which is invariant under ordinary three-dimensional rotation of $x,y,z$.
This is the first clue about the emergence of spacetime from quantum formalism.

On combining independent samples into an ensemble, the vector coordinates $q_x,q_y,q_z$ add and so do their radii $q_0$, which allows the equality (\ref{eq:pure}) to degrade to
\begin{equation}
    \langle q_x \rangle^2 + \langle q_y \rangle^2 + \langle q_z \rangle^2 \le \langle q_0 \rangle^2
\label{eq:mixed}
\end{equation}
With complex coefficients $\phi$, the three generators (\ref{eq:ABC}) define the 6-parameter \emph{Lorentz group} of transformations
\begin{equation}
    \psi' = \exp(\phi_x\Pauli_x + \phi_y\Pauli_y + \phi_z\Pauli_z)\,\psi
\end{equation}
under which
\begin{equation}
 \textrm{$\langle q_0 \rangle^2 - \langle q_x \rangle^2 - \langle q_y \rangle^2 - \langle q_z \rangle^2$ \quad is invariant.}
\label{eq:invariant}
\end{equation}
%

\section{Spin}

Writing the complex coefficients as $\phi = -(\xi + i\eta)/2$, the simplest operation is $\Pauli_z$ with imaginary coefficient $\phi_z = -i\eta/2$, which spins complex  $\psi_\up$ and $\psi_\dn$ in phase.
\begin{equation}
     \pair{{\psi_\up}'}{{\psi_\dn}'} 	= \pair{e^{- i\eta/2} \ \quad 0\ }{\ 0 \ \quad e^{+ i\eta/2}} \pair{\psi_\up}{\psi_\dn}
\label{eq:spin}
\end{equation}
Correspondingly, the observable spins $q$ in (\ref{eq:rates}) and (\ref{eq:quantity}) rotate about $z$ as
\begin{equation}
    \begin{array}{l}
         q_0' = q_0					\\[4pt]
         q_x' = q_x\cos\eta - q_y\sin\eta		\\[4pt]
         q_y' = q_x\sin\eta + q_y\cos\eta	\\[4pt]
         q_z' = q_z     	       				\end{array}
\label{eq:rotate}
\end{equation}
All spins rotate equally, so the observable ensemble rotates undistorted about $z$.

In the context of three-dimensional rotation of $x,y,z$, the ensemble average $\langle q\rangle$ is given symbol $2J$, where $J$ transforms as
\begin{equation}
    \quat{J_0'}{J_x'}{J_y'}{J_z'}	=  \left(\begin{array}{lccr}  \ 1\phantom{x}    	& \ 0             	& 0                 			&  0	\\
											\ 0   				& \ \cos\eta  	& -\sin\eta				&  0 	\\
                                								\ 0   				& \ \sin\eta  	& \phantom{-}\cos\eta        &  0	\\
                                								\ 0   				& \ 0             	& 0                 			& \ \ 1	\end{array}\right) 	\quat{J_0}{J_x}{J_y}{J_z}
\label{eq:rotation}
\end{equation}
with the general invariant (\ref{eq:invariant}) being $J_0^2 - J_x^2 - J_y^2 - J_z^2$.
When, later, space and time have been constructed, $J$ becomes identifiable as angular momentum (including intrinsic spin) in units of~$\hbar$.
The observable invariants under rotation are
\begin{equation}
    J_0 \qquad \textrm{and}\qquad   \surd(J_x^2 + J_y^2 + J_z^2)
\end{equation}

Although rotation is again recognisable in the equations, this is without explicit reference to space which is yet to be constructed.
Observe, though, the commonly-remarked angle-doubling homomorphism between $\eta/2$ which operates on $\psi$ (group SU(2) rotation of a complex plane) and $\eta$ which operates on the observable spin $q$ (group SO(3) usually visualised as rotation of a 3-sphere).
Rotational invariance under SO(3) has emerged automatically from the formalism, and is not imposed as a property of pre-supposed isotropic space \cite{Penrose+Rindler:1984v1,Gull+Lasenby+Doran:1993a}.

\section{Momentum}

We can also apply a real coefficient $\phi_z = -\xi/2$ to $\Pauli_z$.
This rebalances $\psi_\up$ and $\psi_\dn$ as
\begin{equation}
     \pair{{\psi_\up}'}{{\psi_\dn}'} 	= \left(\begin{array}{cc} e^{- \xi/2}	& 0 \\ 0 & e^{+ \xi/2} \end{array}\right) \pair{\psi_\up}{\psi_\dn}
\end{equation}
Correspondingly, the $q$'s in (\ref{eq:rates}) and (\ref{eq:quantity}) transform as
\begin{equation}
    \begin{array}{l}
         q_0' = q_0 \cosh\xi	- q_z\sinh\xi		\\[4pt]
         q_x' = q_x            					\\[4pt]
         q_y' = q_y                	 				\\[4pt]
         q_z' = - q_0\sinh\xi + q_z\cosh\xi     	 \end{array}
\end{equation}
On replacing the ensemble average $\langle q\rangle$ by symbol $p$ for this new set of real properties, we have
\begin{equation}
    \quat{p_0'}{p_x'}{p_y'}{p_z'}	=  \left(\begin{array}{ccccc} \phantom{-}\cosh\xi    	& \ 0       & \phantom{xx} 	& 0   		& -\sinh\xi  	\\
												0				& \ 1       & \ 	& 0   		& 0                		\\
                    										0               		& \ 0       & \ 	& 1   		& 0                 		\\
                    										-\sinh\xi    			& \ 0       & \ 	& 0   		& \phantom{-}\cosh\xi        	\end{array} \right)	\quat{p_0}{p_x}{p_y}{p_z}
\label{eq:boost}
\end{equation}
Instead of rotation, this is a Lorentz boost along $z$, with $p_x$ and $p_y$ unaffected and $p_0^2 - p_z^2$ invariant.
Recognising this from relativistic kinematics, we identify $p_0$ as energy $E$ and $(p_x,p_y,p_z)$ as momentum.
Energy and momentum transform as scalar and vector under rotation, and we write the general invariant (\ref{eq:invariant})
 under boost as
\begin{equation}
	 E^2 - p_x^2 - p_y^2 - p_z^2 = m^2
\label{eq:Eppp}
\end{equation}
where we identify $m$ as (rest-)mass.
We see that mass behaves under transformation (\ref{eq:boost}) as a 4-vector with Minkowski metric ${\tt diag}(1, -1, -1, -1)$.	

Qubits, and consequently all their various combinations, have quantity allied to a directional spin, and energy allied to a directional momentum.
Momentum is imaginary spin and spin is imaginary momentum.
But we do not yet have the time and space within which energy and momentum are conventionally given meaning.

\section{Spacetime}

In the complex environment of quantum theory, generators A and B have yielded a real 3-vector called spin and a dual real 3-vector called momentum.
Each had an associated scalar making 8 properties in all.
Complex $2\mathord\times2$ matrices can support 8 components, but the Hermitian matrices which define observables are only a 4-parameter subset (the Pauli matrices with identity).
Something's missing, and it's the third and final allowed multiplication operator C.

Generator C (equation \ref{eq:phi}C) authorises differentials, thereby allowing integration.
\begin{equation}
	\pair{u'}{U'} =  \mat{1}{0}{t}{1} \pair{u}{U} = \pair{u}{U+ut} \quad\quad \left\{ \begin{array}{l} \delta u = 0 \\ \delta U = u\,\delta t \end{array} \right.
\label{eq:galileo}
\end{equation}
This generates a new parameter $U$ as the integral of an existing parameter $u$ without changing the latter (Galileo's principle).
Conversely, $u$ can be recovered as its differential.

We now inquire into the phase of $\psi$ as in $\psi \propto e^{-i\theta}$  (\ref{eq:complex-number}), with the minus sign being physicists' convention.
For an object of mass $m$, we can define a scaled phase $\tau$ through
\begin{equation}
	d\theta = m\,d\tau
\end{equation}
in which we identify $\theta$ with $U$ in (\ref{eq:galileo}).
Phase cannot be rescaled because it's $2\pi$-periodic, so neither can phase differences $\delta\theta$.
So, with $d\theta$ being invariant to re-orientation and $m$ transforming as a 4-vector under (\ref{eq:Eppp}), it follows that $d\tau$ also transforms as a 4-vector under the same Minkowski metric with invariant
\begin{equation}
    (dt)^2 - (dx)^2  - (dy)^2  - (dz)^2 = (d\tau)^2
\label{eq:dtau}
\end{equation}
having inner product
\begin{equation}
    E\,dt - p_x dx - p_y dy - p_z dz = d\theta
\end{equation}
from which we see that phase $\theta$ is the physicists' \emph{action} \cite{Feynman:1948}.
(In general coordinates, $(dt,dx,dy,dz)$ would be deemed contravariant while $(E,p_x,p_y,p_z)$ would be covariant.)

We now identify $\tau$ as proper time, $t$ as clock time, and $\bfx = (x,y,z)$ as spatial location.
\textbf{This is why our description of space uses three dimensions.}
Velocity defined as $\bfv = d\bfx/dt$ is automatically bounded by 1 (the speed of light) through (\ref{eq:dtau}), so \emph{Einstein's postulate to that effect is not needed} --- it emerges from the formalism.
Meanwhile, our descriptions of time and space inherit continuity from necessarily-continuous phase.
Continuity is not an assumption, it is a requirement.
\textbf{Three-dimensional space and one-dimensional time are identified as integrals of Lorentz boosts.}
Instead of spacetime being a pre-existing framework into which energy and momentum are inserted, it is a construction emerging from those dynamical observables.

The wave-like nature of matter and energy is due to the periodicity of the phase $\phi$ (\ref{eq:complex-number}).
With $\psi \propto e^{-i\theta}$, Schr\"odinger's equation
\begin{equation}
    i\frac{\partial \psi}{\partial t} = E\psi
\end{equation}
and the momentum operators
\begin{equation}
    -i\frac{\partial \psi}{\partial x} = p_x\psi\,,\quad -i\frac{\partial \psi}{\partial y} = p_y\psi\,,\quad -i\frac{\partial \psi}{\partial z} = p_z\psi
\end{equation}
are immediate, with the linearity of Schr\"odinger's equation manifest.
The factors of $i$ ensure that the operators for energy and momentum are Hermitian, so those properties are observable in accordance with experience.

In summary, relativistic kinematics, with three dimensions of space and one of time, emerges automatically from the foundational structure of quantum theory.
This language of physics, in the form of the 6-parameter Lorentz group of spin and momentum, augmented by 4-parameter spacetime,
  follows inevitably from the necessity of introducing uncertainty at the outset of our inquiry.
As macroscopic beings, we are at the end of this chain of development.
\begin{equation*}
    \boxed{\boxed{
    \begin{array}{l}
            \textbf{Uncertainty}																										\\
                   \qquad\qquad								\displaystyle 	\ \mathop{\Longrightarrow}^{\rm A}\				\textrm{complex numbers}		\\
                   \qquad\qquad\qquad\qquad	 				\displaystyle 	\ \mathop{\Longrightarrow}^{\rm B}\				\textrm{spin \& momentum}	\\
                   \qquad\qquad\qquad\qquad\qquad\qquad 		\displaystyle 	\ \mathop{\Longrightarrow}^{\rm C}_{\phantom{X}}\	\textrm{spacetime}			\\
                   \qquad\qquad\qquad\qquad\qquad\qquad\qquad\qquad 			\ \mathop{\Longrightarrow}\ 					\textbf{Perception\!}			\\
	\end{array}
        }}
\end{equation*}

Not only do the elementary symmetries of associative commutativity and associative distributivity force the mathematics used to quantify objects and events, but they also force the descriptions underlying our perception of objects and events \cite{Pellionisz+Llinas:1982}.
These inherent perceptual descriptions lead one to feel that space and time are an objective reality, whereas our analysis shows that three-dimensional space and one-dimensional time are the only consistent description one can have of reality.
Perhaps that's why it has taken centuries for this structure to be uncovered.

It will not escape notice that spatial location is accumulated as the time-integral of local velocity.
Consequently, the curvature of space, if any, is not defined in the language, but is part of the content of physics, along with basic parameters of particle physics and cosmology.
In this paper, we have addressed the language only, leaving content for other investigation.

\section{Conclusions}

We have argued that objects must be represented, not by classical scalars, but by pairs of numbers fusing quantity and uncertainty, and have shown how they can be so quantified.
Associative commutativity of combination and partition results in the component-wise sum rule, and associative distributivity of chaining or sequencing results in three possible product rules.
These define the arithmetical rules to which our formalism must adhere.
Scalars have only one product rule but pairs have three.

Only one of the three product rules (A) yields proper (probabilistic) predictions.
Applied to pairs, it imposes complex quantum amplitudes controlled by the Feynman sum and product rules with predictions given by the Born rule.
By construction, quantum formalism is thus engineered as a probabilistic theory incorporating uncertainty.
Uncertainty appears as phase and the theory is engineered to predict objects' behavior in the form of likelihood functions, which is \emph{physics}.
Given those likelihoods, Bayesian analysis then computes posterior probabilities, which assess the models in the light of outcomes as actually observed, which is \emph{inference}.
In this way physics and inference are distinct, but mutually consistent, with physics making predictions about the world and inference using those predictions along with experimental outcomes to learn about the world.
Of course, there can be no conflict between standard classical probability and quantum applications because the sum and product rules for scalars and pairs follow together from common foundational symmetries to give consistent arithmetical rules.

Every independent component of a compound object has its own unknown phase in a wide-ranging ensemble of possibilities.
For large objects, such phases average out leaving classically scalar macroscopic quantities.
Where components are not independent because of related construction, their connected phases are said to be ``entangled''.
Entanglement yields extra information about an object, inaccessible to classical scalar inquiry as exemplified by the
Bell inequalities \cite{Bell:1964,Werner+Wolf:2001}.

Beyond phase and entanglement, which are underpinned by product rule A, lie rules B and C which take the scope further.
Those rules must also play a part --- otherwise some additional assumption would be required to prevent it.
Rule B generates the Pauli matrices, which yield spin, energy, and momentum.
Rule C integrates those to construct 3+1-dimensional relativistic spacetime.
This explains why space has three dimensions, and why Einstein's postulate of a speed limit (that of light) was correct \cite{Einstein:1905}.

Upon acknowledging that quantification carries uncertainty with it, we find that quantum formalism and relativistic spacetime both follow without further assumption.
The rules follow from logical derivation, and not from experimentation which must and does conform.
As first demonstrated by  \cite{Cox:1946}, robust formalism is not invented or discovered, but is openly engineered to conform to fundamental symmetries.
There is then no mystery about why the resulting mathematics works.
It has to.

We do  not, of course, make the philosophically contentious claim to have proved our findings in some absolute sense.
All we claim is that any hypothetical alternative theory not conforming to standard quantum formalism and standard spacetime must be incompatible with the foundational symmetries, so that combination and/or sequencing must go awry in that world.
In principle, a world could be so deeply interconnected that separating individual objects was impossible.
In practice, ours isn't, so we stand by our findings as a uniquely consistent mathematical description of the world we live in.

\medskip
\noindent{\underline{Scalars}:}
$$
\begin{array}{|l|}
\hline
	\tall{	 \qquad \begin{array}{ll} 	\textrm{{\bf Shuffling}} &:= \textrm{commutative + associative} 	\\
		 									\textrm{{\bf Sequencing}} &:= \textrm{distributive\ \ +\ \;associative} 	\end{array}						}
\\ \hline
	\tall{	\left. \begin{array}{lcl} 	\textrm{Shuffling}\ \ \ \ \ \ \  		& \rightarrow	&	\textrm{Sum rule, quantification}		\end{array} \right\} \textbf{ Measures}	}
\\ \hline
	\tall{	\left. \begin{array}{lcl} 	\textrm{Shuffling} 		& \rightarrow	&	\textrm{Sum rule, quantity} 			\\
							\textrm{Sequencing} 		& \rightarrow	&  	\textrm{Product rule, proportion} 		\end{array} \right\} \textbf{Probability}			}
\\ \hline
\end{array}
$$
\medskip

\noindent{\underline{Pairs}:}
$$
\begin{array}{|c|}
\hline
	\tall{	\begin{array}{c} \textrm{Quantity \& uncertainty} \ \rightarrow\ \textrm{represent by number {\bf pair}} \end{array}						}
\\ \hline
	\tall{	\begin{array}{lcl} 	\textrm{Shuffling} 	& \rightarrow	&	\textrm{component-wise {\bf sum rule}}			\\
						\textrm{Sequencing}	& \rightarrow	&	\textrm{three allowed {\bf product rules} A,B,C}		\\
										& 			&	\textrm{\quad A = complex multiplication}			\\
										& 			&	\textrm{\quad B = hyperbolic variant}				\\
										& 			&	\textrm{\quad C = integrator}					\end{array}			}
\\ \hline
	\begin{array}{l}	\textrm{1 property (existence)} \rightarrow \\
				\quad\qquad\left\{	\begin{array}{l} 	
                                                     			\textrm{Only A is consistent with probability}								\\
							\therefore\ \textrm{pair = {\bf complex number} $\psi$, with}							\\
							\quad	\textrm{ complex $+$ and $\times$ (     }	\textbf{Feynman picture}	\textrm{)}		\\
							\quad	\textrm{ and quantity = $|\psi|^2$ (the }	\textbf{Born rule}		\textrm{)}		\end{array} \right. \end{array}	
\\ \hline
	\begin{array}{l}		\textrm{2 properties (qubit)} \rightarrow \\
				 \quad\left\{ \begin{array}{l} 	\textrm{A,B,C can all apply}											\\
									\textrm{A,B $\rightarrow$ Pauli matrices $\rightarrow$ {\bf spin}, {\bf 4-momentum}}	\\
									\textrm{C on 4-momentum $\rightarrow$ {\bf relativistic spacetime}}				\end{array} \right.	\end{array}
\\ \hline
\end{array}
$$


\section*{Acknowledgments}

The authors thank Hans Eggers, Ariel Caticha, Seth Chaiken, Keith Earle, Cecilia Levy, Oleg Lunin, Matthew Szydagis, Steve Gull, Anton Garrett, and Chris Doran
 --- also Dylan VanAllen and the Spring 2020 Advanced Physics Laboratory (APHY 335Z) and Fall 2020 Bayesian Data Analysis  (APHY/ICSI/ IINF 451/551) classes at the University at Albany
  --- for their helpful discussions on the material related to and presented in this paper.
Knuth also thanks the Foundational Questions Institute (FQXi) for their Essay Contests, which have encouraged many of us physicists to carefully think and write about these deep issues.

\section*{Appendix}

This analysis updates and clarifies material originally presented in Appendix A of the GKS paper \cite{GKS:PRA}.
In that paper, the primes in its equation (19) were inconsequentially misplaced --- they should have been on its right-hand $\g$'s, not the left.

\noindent \underline{\bf Associative commutativity} of $+$  requires combination $\bfc = \bfa + \bfb$ of $n$-tuples to be represented --- up to isomorphism
 --- by component-wise addition (the \emph{sum rule}, which is unique for any $n$).
\begin{equation*}
\begin{array}{r}
	\left. \begin{array}{cl} \bfx + \bfy = \bfy + \bfx			& \textrm{(commutative)}	\\
					(\bfx+\bfy)+\bfz = \bfx + (\bfy + \bfz)	& \textrm{(associative)} 	\end{array}\right\}	\quad\forall\, \bfx,\bfy,\bfz \phantom{xxxxx}		\\
	 \Longrightarrow\quad c_i = a_i + b_i \quad\forall\,i																					\end{array}
\end{equation*}
This linear form is invariant to non-singular linear transformation of axes.
Distributivity of $\,\cdot\,$ over $+$ then requires $n$-tuple connection $\bfc = \bfa\cdot\bfb$ to be bilinear multiplication defined by $n^3$ coefficients $\g$.
\begin{equation*}
\begin{array}{r}
	\left. \begin{array}{cl} \bfz\cdot(\bfx\mathord+\bfy) = \bfz\cdot\bfx + \bfz\cdot\bfy				& \textrm{(left distributive)}	\\
					(\bfx\mathord+\bfy)\cdot\bfz = \bfx\cdot\bfz + \bfy\cdot\bfz	& \textrm{(right distributive)} 	\end{array}\right\}	\quad\forall\, \bfx,\bfy,\bfz \phantom{xxxx}		\\
	\displaystyle \Longrightarrow\quad c_i = \sum_{j=1}^n \sum_{k=1}^n \g_{ijk} a_j b_k \quad\forall\,i																\end{array}
\end{equation*}
\underline{\bf Associative distributivity} imposes $n^4$ quadratic constr\-aints (not all independent) on those $n^3$ $\g$'s.
\begin{equation*}
\begin{array}{r}
	(\bfx\cdot\bfy)\cdot\bfz = \bfx\cdot(\bfy\cdot\bfz)			\ \  \textrm{(associative distributivity)}	\quad\forall\, \bfx,\bfy,\bfz \phantom{xx}		\\
	\displaystyle \Longrightarrow\qquad \sum_{t=1}^n (\g_{tpq} \g_{rst} - \g_{tsp} \g_{rtq}	) = 0 \quad \forall\, p,q,r,s					\end{array}
\end{equation*}
The associativity is to be nontrivial: $\bfx \cdot \bfy \cdot\bfz \not\equiv {\bf 0}$.
Associative commutativity and associative distributivity are our foundational symmetries.
The constraints are simple quadratics, which accords with intuition that generality should be simple.

We seek standardised values for the multiplication coefficients $\g$ that we can adopt as agreed conventions for representing families connected by non-singular transformation of axes.
Each such standard will be called a \emph{product rule}.

For a start, we can scale axes by	
$$
	x_j \quad\longrightarrow\quad x_j' = x_j / \lambda \qquad\Longrightarrow\qquad \g_{jjj}' = \lambda \g_{jjj}
$$
for each index $j$ separately.
This lets us set the coefficients $\g_{jjj}$ (all 3 suffices the same) to 0 or $+1$ for each $j$.
\subsection*{One dimension $n=1$}

With only one coefficient $\g_{111}$, setting it to 0  would be trivial and useless so we assign
$$
	\g_{111} = 1
$$
Alternative conventions are possible, as when proportions are presented as percentages ($\g_{111} = 1/100$), but the unit assignment is standard.
In one dimension, distributivity allows  \emph{only one product rule} (which is ordinary scalar multiplication) and associativity follows trivially.
But in higher dimension more than one product rule can accompany the unique sum rule.

\subsection*{Two dimensions $n=2$}

There are now $2^3=8$ $\g$'s, which we lay out as an array
$$
	\bfgamma = \left[\begin{array}{cccc}	\g_{111}	& \ \g_{112}	& \ \g_{121}	&\  \g_{122}		\\
								\g_{211}	& \ \g_{212}	& \ \g_{221}	&\  \g_{222}		\end{array}\right]
$$
The $2^4=16$ associativity constraints are
$$
	\begin{array}{c}
\g_{111} \g_{111} 	- \g_{111} \g_{111} 	+ \g_{211} \g_{112} 	- \g_{211} \g_{121}   	= 0 \\
\g_{111} \g_{121} 	- \g_{121} \g_{111} 	+ \g_{211} \g_{122} 	- \g_{221} \g_{121}   	= 0 \\
\g_{111} \g_{211} 	- \g_{111} \g_{211} 	+ \g_{211} \g_{212} 	- \g_{211} \g_{221}   	= 0 \\
\g_{111} \g_{221} 	- \g_{121} \g_{211} 	+ \g_{211} \g_{222} 	- \g_{221} \g_{221}   	= 0 \\
\g_{112} \g_{111} 	- \g_{111} \g_{112} 	+ \g_{212} \g_{112} 	- \g_{211} \g_{122}   	= 0 \\
\g_{112} \g_{121} 	- \g_{121} \g_{112} 	+ \g_{212} \g_{122} 	- \g_{221} \g_{122}   	= 0 \\
\g_{112} \g_{211} 	- \g_{111} \g_{212} 	+ \g_{212} \g_{212} 	- \g_{211} \g_{222}   	= 0 \\
\g_{112} \g_{221} 	- \g_{121} \g_{212} 	+ \g_{212} \g_{222} 	- \g_{221} \g_{222}   	= 0 \\
\g_{121} \g_{111} 	- \g_{112} \g_{111} 	+ \g_{221} \g_{112} 	- \g_{212} \g_{121}   	= 0 \\
\g_{121} \g_{121} 	- \g_{122} \g_{111} 	+ \g_{221} \g_{122}	- \g_{222} \g_{121}   	= 0 \\
\g_{121} \g_{211} 	- \g_{112} \g_{211} 	+ \g_{221} \g_{212} 	- \g_{212} \g_{221}   	= 0 \\
\g_{121} \g_{221} 	- \g_{122} \g_{211} 	+ \g_{221} \g_{222} 	- \g_{222} \g_{221}   	= 0 \\
\g_{122} \g_{111} 	- \g_{112} \g_{112} 	+ \g_{222} \g_{112} 	- \g_{212} \g_{122}   	= 0 \\
\g_{122} \g_{121} 	- \g_{122} \g_{112} 	+ \g_{222} \g_{122} 	- \g_{222} \g_{122}   	= 0 \\
\g_{122} \g_{211} 	- \g_{112} \g_{212} 	+ \g_{222} \g_{212} 	- \g_{212} \g_{222}   	= 0 \\
\g_{122} \g_{221} 	- \g_{122} \g_{212} 	+ \g_{222} \g_{222} 	- \g_{222} \g_{222}   	= 0 \\
	\end{array}
	\eqno{\begin{array}{r} (1) \\ (2) \\ (3) \\ (4) \\ (5) \\ (6) \\ (7) \\ (8) \\ (9) \\ (10) \\ (11) \\ (12) \\ (13) \\ (14) \\ (15) \\ (16) \end{array}}
$$

As well as scaling, we can also shear one of the axes by the other, as in
$$x_2 \longrightarrow x_2' = x_2 - \alpha x_1.$$
The $\g$'s then transform with, in particular,
\begin{equation*}
\begin{split}
	\g_{211}' = \g_{211} &+ (\g_{221} + \g_{212} - \g_{111})\alpha \\&+ (\g_{222} - \g_{121} - \g_{112})\alpha^2 - \g_{122}\alpha^3
\end{split}
\end{equation*}
Full-rank cubic equations have at least one real root, enabling us to set $\g_{211}=0$ unless $\g_{122}= 0$.
Finally, we can interchange axis labels $1 \rightleftharpoons 2$, so consequently we can always select \underline{$\g_{211}=0$}.

We now choose $\g_{111}$ to be either 1 or~0.

\subsubsection*{Two dimensions $n=2$, Choice \#1.}

On appending \underline{$\g_{111}=1$} to the original \underline{$\g_{211}=0$}, the associativity equations reduce to
$$
	\begin{array}{c}
  		\g_{121} \g_{221} = 0 								\\
  		\g_{112} \g_{212}  	= 0								\\
  		\g_{221}(\g_{221}-1)  = 0								\\
  		\g_{212}(\g_{212}-1)  = 0								\\
  		\g_{122}(\g_{212} - \g_{221}) = 0						\\
 		 \g_{122}(\g_{121} - \g_{112})  = 0						\\
  		\g_{112}(\g_{221}-1) = \g_{121}(\g_{212}-1)				\\
  		\g_{122}(\g_{221}-1) = \g_{121}(\g_{222}-\g_{121}) 			\\
  		\g_{122}(\g_{212}-1) = \g_{112}(\g_{222} -\g_{112}) 			\\
  		\g_{212}(\g_{222}-\g_{121}) = \g_{221}(\g_{222}-\g_{112}) 	\\
	\end{array}
	\eqno{\begin{array}{r} (2') \\ (5') \\ (4') \\ (7') \\ (6') \\ (14') \\ (9') \\ (10') \\ (13') \\ (8') \end{array}}
$$
From $(4')$, $\g_{221}=1$ or $0$; and from $(7')$, $\g_{212}=1$ or $0$.
This allows four choices (\#$1\mathord\cdot 1$, \#$1\mathord\cdot 2$, \#$1\mathord\cdot 3$, \#$1\mathord\cdot 4$).

\subsubsection*{Two dimensions $n=2$, Choice \#$1\mathord\cdot 1$.}

Choice \#$1\mathord\cdot 1$ adds \underline{$\g_{221} = 1$ and $\g_{212} = 1$} to $\g_{111}=1$ and $\g_{211}=0$.

From $(2')$, $\g_{121}=0$; and from $(5')$, $\g_{112}=0$.
All equations are then satisfied, with $\g$ taking the form
$$
	\bfgamma = \left[\begin{array}{cccc}	  1		& \  0		& \ 0		& \ \theta		\\		0 	& 	\ 1	& 	\ 1	& 	\ \phi		\end{array}\right]	
$$
Free parameter $\phi$ can be sheared away by  $x_1' = x_1 + \frac{1}{2} \phi x_2$, giving
$$
	\bfgamma = \left[\begin{array}{cccc}	  1		& \ 0		& \ 0		& \ \theta'		\\		0 	& 	\ 1	& 	\ 1	& 	\ 0		\end{array}\right]	
$$
On rescaling the ``2'' axis, $\g_{122}' = \theta'$ can be set to standard values of $-1$ (if originally negative) or 1 (if originally positive) or 0 (if originally zero).
This gives three standard forms, similar in style but not related by any real coordinate transformation:
$$
	\bfgamma = \left[\begin{array}{cccc}	  1		& \ 0		& \ 0		& \ \minus 1\!	\\		0 	& 	\ 1	& 	\ 1	& 	\ 0		\end{array}\right]	
\textrm{\ and\ }
	 		\left[\begin{array}{cccc}	  1		& \ 0		& \ 0		& \ 1			\\		0 	& 	\ 1	& 	\ 1	& 	\ 0		\end{array}\right]	
\textrm{\ and\ }
	 		\left[\begin{array}{cccc}	  1		&\ 0		& \ 0		& \ 0			\\		0 	& 	\ 1	& 	\ 1	& 	\ 0		\end{array}\right]	\eqno{\textrm{(A,B,C)}}
$$

\subsubsection*{Two dimensions $n=2$, Choice \#$1\mathord\cdot 2$.}
Choice \#$1\mathord\cdot 2$ adds \underline{$\g_{221} = 0$ and $\g_{212} = 1$} to $\g_{111}=1$ and $\g_{211}=0$.

From $(5')$, $\g_{112}=0$; from $(6')$, $\g_{122}=0$; and from $(8')$, $\g_{121} = \g_{222}$.
All equations are then satisfied, with $\g$ taking the form
$$
	\bfgamma = \left[\begin{array}{cccc}	  1		& \ 0		& \ \phi	& \ 0			\\		0 	& 	\ 1	& 	\ 0	& 	\ \phi		\end{array}\right]	
$$
Free parameter $\phi$ can be sheared away by  $x_1' = x_1 + \phi x_2$ to give the standard form
$$
	\bfgamma = \left[\begin{array}{cccc}	  1		& \ 0		& \ 0		& \ 0			\\		0 	& 	\ 1	& 	\ 0	& 	\ 0		\end{array}\right]	\eqno{\textrm{(D)}}
$$

\subsubsection*{Two dimensions $n=2$, Choice \#$1\mathord\cdot 3$.}
Choice \#$1\mathord\cdot 3$ adds \underline{$\g_{221} = 1$ and $\g_{212} = 0$} to $\g_{111}=1$ and $\g_{211}=0$.

This is the $\bfa\cdot\bfb \rightleftharpoons \bfb\cdot\bfa$ reversal $\g_{ijk} \rightleftharpoons \g_{ikj}$ of choice \#$1\mathord\cdot 2$ and leads to the reversed standard form
$$
	\bfgamma = \left[\begin{array}{cccc}	  1		& \ 0		& \ 0		& \ 0			\\		0 	& 	\ 0	& 	\ 1	& 	\ 0		\end{array}\right]	\eqno{\textrm{(E)}}
$$

\subsubsection*{Two dimensions $n=2$, Choice \#$1\mathord\cdot 4$.}
Choice \#$1\mathord\cdot 4$ adds \underline{$\g_{221} = 0$ and $\g_{212} = 0$} to $\g_{111}=1$ and $\g_{211}=0$.

\noindent\phantom{xx.}From $(9')$, $\g_{121} = \g_{112}$; and from $(10')$, $\g_{122} = \g_{121}(\g_{121} - \g_{222})$.
All equations are then satisfied, with $\g$ taking the form
$$
	\bfgamma = \left[\begin{array}{cccc}		1	& \ \ \theta	& \ \ \theta	& \ \theta(\theta\mathord-\phi)	\\		0 	& 	\ \ 0	& 	\ \ 0	& 	\ \ \phi		\end{array}\right]
$$
Free parameter $\theta$ can be sheared away by  $x_1' = x_1 + \theta x_2$ to give
$$
	\bfgamma = \left[\begin{array}{llll}		1	& \ 0		& \ 0		& \ 0			\\		0 	& 	\ 0	& 	\ 0	& 	\ \phi		\end{array}\right]
$$
Scaling $\g_{222} = \phi$ to 0 or 1 then appears to give two standard forms
$$
	\bfgamma = 	\left[\begin{array}{llll}		1	& \ 0		& \ 0		& \ 0			\\		0 	& 	\ 0	& 	\ 0	& 	\ 0		\end{array}\right]	
\textrm{\ and\ }
	  			\left[\begin{array}{llll}		1	& \ 0		& \ 0		& \ 0			\\		0 	& 	\ 0	& 	\ 0	& 	\ 1		\end{array}\right]		\eqno{\textrm{(F,\,G})}
$$
However, form G shears under $(x_1',x_2') = \frac{1}{2}(x_1\mathord-x_2, x_1\mathord+x_2)$ to  form B above, so is not new.

\subsubsection*{Two dimensions $n=2$, Choice \#2.}

On adding \underline{$\g_{111}=0$} to the original \underline{$\g_{211}=0$}, the associativity equations reduce to
$$
	\begin{array}{c}
  		\g_{121} \g_{221}   = 0								\\
  		\g_{112} \g_{212}  = 0								\\
  		\g_{221} \g_{221}  = 0								\\
  		\g_{212} \g_{212}  = 0								\\
  		\g_{112} \g_{221} = \g_{121} \g_{212} 					\\
  		\g_{122}(\g_{212} - \g_{221})  = 0						\\
  		\g_{122}(\g_{121} - \g_{112})  = 0						\\
  		\g_{121}(\g_{222} - \g_{121})  = \g_{122} \g_{221} 			\\
  		\g_{112}(\g_{222} - \g_{112})  =  \g_{122} \g_{212}			\\
 		\g_{221} (\g_{112}  - \g_{222})  = \g_{212} (\g_{121} - \g_{222}) 	\\
	\end{array}
	\eqno{\begin{array}{r} (2'') \\ (5'') \\ (4'') \\ (7'') \\ (9'') \\ (6'') \\ (14'') \\ (10'') \\ (13'') \\ (8'')  \end{array}}
$$
From $(4'')$, $\g_{221}=0$; and from $(7'')$, $\g_{212}=0$; leaving only
$$
	\begin{array}{c}
  		\g_{122}(\g_{121} - \g_{112}) = 0  		\\
  		\g_{121}(\g_{121} - \g_{222}) = 0  		\\
  		\g_{112}(\g_{112} - \g_{222}) = 0  		\\
	\end{array}
	\eqno{\begin{array}{r} (14''') \\ (10''') \\ (13''') \end{array}}
$$
From $(10'')$, $\g_{121}$ is 0 or $\g_{222}$; and from $(13'')$, $\g_{112}$ is 0 or $\g_{222}$; allowing four forms
\begin{equation*}
\begin{split}
	\bfgamma = 	\left[\begin{array}{cccc}	  0	& \ \theta	& \ \theta	& \ \phi	\\	0	& 	\ 0	& 	\ 0	& 	\ \theta	\end{array}\right]	
&\textrm{\ and\ }
				\left[\begin{array}{cccc}	  0	& \ 0		& \ 0		& \ \phi	\\	0	& 	\ 0	& 	\ 0	& 	\ \theta	\end{array}\right]	
\\&\textrm{\ and\ }
	 			\left[\begin{array}{cccc}	  0	& \ \theta	& \ 0		& \ \phi	\\	0	& 	\ 0	& 	\ 0	& 	\ \theta	\end{array}\right]	
\textrm{\ and\ }
	 			\left[\begin{array}{cccc}	  0	& \ 0		& \ \theta	& \ \phi	\\	0	& 	\ 0	& 	\ 0	& 	\ \theta	\end{array}\right]	
\end{split}
\end{equation*}
To be nontrivial, $\bfx \mathord\cdot \bfy\mathord\cdot\bfz \not\equiv {\bf 0}$, these forms need $\g_{222} = \theta \ne 0$.
Hence $\theta$ must be scalable to $+1$.
Equation $(14''')$ then requires $\phi=0$ in the last two forms, which reproduce earlier solutions D and E in interchanged $1\rightleftharpoons2$ form, so offer nothing new.
Meanwhile $\phi$ can be sheared to 0 in the first two forms by $x_1' = x_1 \pm \phi x_2$ (with $+$ sign for the first, $-$ for the second).
Again, these reproduce earlier solutions C and F in interchanged form, so offer nothing new.

\subsection*{Summary}

In two dimensions, associative distributivity allows \emph{six standard product rules}, all different, which we label alphabetically.
$$
	\boxed{
	\bfgamma = \left[\begin{array}{cccc}	  1		& \ 0		& \ 0		& \ \minus 1\!	\\		0 	& 	\ 1	& 	\ 1	& 	\ 0		\end{array}\right]	}
	\,,\qquad	\binom{a_1}{a_2} \cdot \binom{b_1}{b_2} = \binom{a_1b_1 - a_2b_2}{a_1b_2 + a_2b_1}
	\eqno{\textrm{(A)}}
$$
$$
	\boxed{
	\bfgamma = \left[\begin{array}{cccc}	  1		& \ 0		& \ 0		& \ 1			\\		0 	& 	\ 1	& 	\ 1	& 	\ 0		\end{array}\right]	}
	\,,\qquad	\binom{a_1}{a_2} \cdot \binom{b_1}{b_2} = \binom{a_1b_1 + a_2b_2}{a_1b_2 + a_2b_1}
	\eqno{\textrm{(B)}}
$$
$$
	\boxed{
	\bfgamma = \left[\begin{array}{cccc}	  1		& \ 0		& \ 0		& \ 0			\\		0 	& 	\ 1	& 	\ 1	& 	\ 0		\end{array}\right]	}
	\,,\qquad	\binom{a_1}{a_2} \cdot \binom{b_1}{b_2} = \binom{a_1b_1}{a_1b_2 + a_2b_1}
	\eqno{\textrm{(C)}}
$$
$$
 	\boxed{
	\bfgamma = \left[\begin{array}{cccc}	  1		& \ 0		& \ 0		& \ 0			\\		0 	& 	\ 1	& 	\ 0	& 	\ 0		\end{array}\right]	}	
	\,,\qquad\qquad	\binom{a_1}{a_2} \cdot \binom{b_1}{b_2} = a_1\binom{b_1}{b_2}\quad
	\eqno{\textrm{(D)}}
$$
$$
 	\boxed{
	\bfgamma = \left[\begin{array}{cccc}	  1		& \ 0		& \ 0		& \ 0			\\		0 	& 	\ 0	& 	\ 1	& 	\ 0		\end{array}\right]	}	
	\,,\qquad\qquad	\binom{a_1}{a_2} \cdot \binom{b_1}{b_2} = \binom{a_1}{a_2}b_1\quad
	\eqno{\textrm{(E)}}
$$
$$
	\boxed{
	\bfgamma = \left[\begin{array}{llll}		1	& \ 0		& \ 0		& \ 0			\\		0 	& 	\ 0	& 	\ 0	& 	\ 0		\end{array}\right]	}
	\,,\qquad\qquad	\binom{a_1}{a_2} \cdot \binom{b_1}{b_2} = \binom{a_1b_1}{0}\quad
	\eqno{\textrm{(F)}}
$$
Only the first three offer the nondegenerate ``$\textrm{Pair} \cdot \textrm{Pair} \rightarrow \textrm{Pair}$'' product laws that we seek to engineer.
Rules D and E introduce multiplication by scalars, and F confirms ordinary multiplication of those scalars.
Thus the full structure of an associative algebra follows from pair-wise associative commutativity and nontrivial associative distributivity alone, with scalars emerging automatically  as pairs of form $(x,0)$.

\end{document}